# Cataclysm no more:
# New views on the timing and delivery of lunar impactors[1]

Nicolle E. B. Zellner
Department of Physics, Albion College
611 E. Porter St., Albion, MI 49224
nzellner@albion.edu

**Abstract:**

If properly interpreted, the impact record of the Moon, Earth's nearest neighbour, can be used to gain insights into how the Earth has been influenced by impacting events since its formation ~4.5 billion years (Ga) ago. However, the nature and timing of the lunar impactors – and indeed the lunar impact record itself – are not well understood. Of particular interest are the ages of lunar impact basins and what they tell us about the proposed "lunar cataclysm" and/or the late heavy bombardment (LHB), and how this impact episode may have affected early life on Earth or other planets. Investigations of the lunar impactor population over time have been undertaken and include analyses of orbital data and images; lunar, terrestrial, and other planetary sample data; and dynamical modelling. Here, the existing information regarding the nature of the lunar impact record is reviewed and new interpretations are presented. Importantly, it is demonstrated that most evidence supports a prolonged lunar (and thus, terrestrial) bombardment from ~4.2 to 3.4 Ga and not a cataclysmic spike at ~3.9 Ga. Implications for the conditions required for the origin of life are addressed.

---

[1] NEBZ fondly acknowledges and remembers Dr. Jim Ferris, who directed one of the first Astrobiology Institutes, the NASA Specialized Center of Research and Training (NSCORT) for the Study of Origins of Life, at Rensselaer Polytechnic Institute (Troy, NY), where she started her career in this field.

# 1. Introduction

As the nearest planetary neighbour to Earth, and one that has no atmosphere, no plate tectonics, and virtually no water, the Moon preserves a record of impacts since its formation 4.53 billion years ago (Ga). Aside from impacts, then, the Moon's surface is undisturbed by geological processes common to Earth. Consequently, the nature of the lunar impact flux has been a topic of enduring concern for the planetary science and astrobiology communities. For the lunar community, refined constraints on the timing and duration of the lunar impact flux can lead to a better understanding of how the lunar surface has evolved, and in particular, the timing of the appearance of the large nearside impact basins and the extent of their ejecta. Importantly, the timing of the delivery of those impactors allows for improved understanding of the evolution of the Solar System in general. In particular, critical questions to be answered include (1) determining the form of the large-impact distribution with respect to time (e.g., smooth decline versus cataclysmic spike), (2) whether there is periodicity in Earth-Moon cratering history (e.g., Alvarez and Muller, 1984), and (3) the applicability of the lunar record to other planets. Additionally, and of relevance to the astrobiology community, the lunar impact record serves as a touchstone on which impactor flux and distribution throughout the Solar System, and in particular, on a young potentially habitable Earth (or other Earth-like planet), are based. Of great interest to astrobiology and the study of the origin of life is the impact flux prior to ~3.7 Ga ago, and specifically, whether or not early life, if it existed on Earth before 4 Ga ago, may have been destroyed (Maher and Stevenson, 1988; Sleep et al. 1989; Ryder, 2002; Abramov and Mojzsis, 2009) during these early impact events.

Ages of lunar samples (many of which have unknown provenances) brought back by the Apollo astronauts and uncrewed Luna missions, ages of lunar meteorites with unknown locations of origins, stratigraphy (e.g., ejecta emplacement) of the large impact basins and smaller craters, and crater counting (e.g., the number of secondary craters in the ejecta blankets, the number of small craters in the basins) can all serve as methods by which these issues can be addressed. Since the arrival of Apollo and Luna mission samples to the present time, interpretations of lunar data, samples, and geology have changed, and with these changes, a new understanding of the lunar impact flux has emerged, one that has profound implications for the conditions under which early terrestrial life may have formed and survived.

## 2. Initial Interpretations: Orbital Images and Apollo Impact Melt Samples

Early ideas regarding the lunar impact flux were proposed as a result of studies of the lunar surface. Based on orbital images that showed lightly cratered maria (i.e., the large dark areas filling the large basins) and crater-saturated highlands (Figure 1), Baldwin (1949, 1964) and Hartmann (1965, 1966) suggested that the Moon's early cratering rate was roughly 200× the average post-major-mare extrusion rate (i.e., after 3.3 Ga, as determined by Apollo samples) and that the peak rate was probably much higher. Over time, this rate declined monotonically ("smooth decline" in Figure 2), as the larger impacting objects were swept up by planets and moons, or tossed out of the Solar System altogether, due to gravitational interactions.

Once the Apollo samples were brought to Earth and analysed however, another scenario emerged when no lunar impactites (of 18 analysed) showed U-Pb ages older than ~3.9 Ga ("cataclysm" in Figure 2). One scenario proposed by Tera et al. (1974) and others suggested that the Moon experienced episodic impacts (see Fig. 6 in Tera et al., 1974), the last of which occurred at ~3.9 Ga, with the Imbrium, Crisium, and Orientale Basins (Figure 1), and perhaps others, all forming at this time within 30 million years (My) of each other. This final episode is known as the "terminal lunar cataclysm" (Turner et al. 1973; Tera et al. 1974). In an extreme interpretation of these early Apollo and Luna data, Ryder (1990) proposed that the Moon experienced only light bombardment in its first 600 Ma (i.e., essentially no impacts occurred before 3.9 Ga) and that all of the large near-side basins formed in a narrow window of time centered around 3.85 Ga. Thus, Ryder (1990) reinforced "View 1" of Tera et al. (1974), and it became the prevalent view in the scientific community. Later analyses of Apollo 15 and 17 impact melts (Ryder and Dalrymple, 1993, 1996), which did not show ages >3.9 Ga, appeared to support this extreme interpretation of the lunar cataclysm (Figure 1, 2$^{nd}$ column in Table 1). Other investigators (e.g., Tera et al. 1974 (View 2); Neukum et al. 1975b; Hartmann et al. 2000), however, in order to explain the saturated crater record in the lunar highlands, proposed that the Moon must have experienced some impacts prior to 3.85 Ga. Thus, an "early intense bombardment" (e.g., Hartmann et al. 2000) may have occurred (Figure 1). South Pole-Aitken Basin, recognized as the largest and oldest impact basin on the Moon (e.g., Stuart-Alexander 1978; Wilhelms 1979; Wilhelms 1987), has been assigned a highly uncertain age of ~4.3 – 4.05 Ga (pre-Nectarian; e.g., Wilhelms 1987).

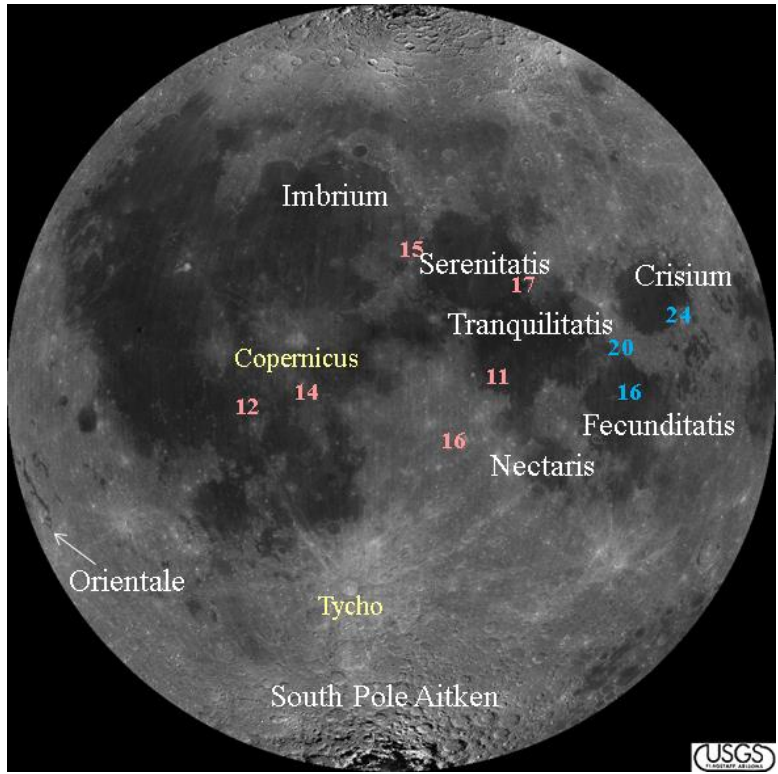

**Figure 1.** Image of the Moon, with impact basins (white) and prominent craters (yellow) noted. Large dark areas are the maria and lighter areas are the highlands. Apollo (pink) and Luna (blue) sites have also been noted.

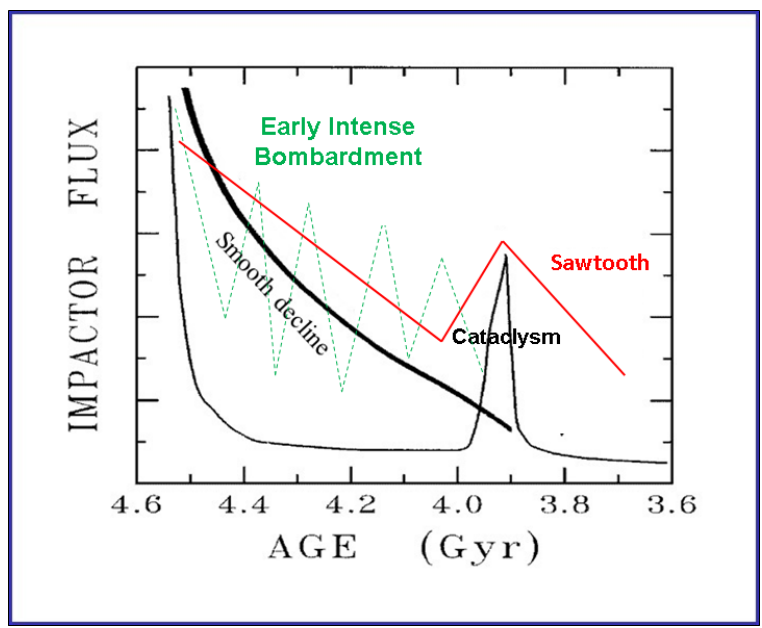

**Figure 2.** Summary of lunar impact scenarios, as presented in the text. Graph is by J. W. Delano and modified from Zellner (2001). General flux curves are interpretations of evidence presented in Baldwin (1949, 1964), Hartmann (1965, 1966, 2000), Tera et al. (1974), Neukum et al. (1975b, 2001), Neukum and Ivanov (1994), and Morbidelli et al. (2012).

It is noted that the term "cataclysm" (or "spike") has sometimes been used interchangeably (e.g., Ćuk et al. 2010; Abramov et al. 2013; Werner 2014) with "late heavy bombardment" (LHB), the influx of material that was thought to have happened in the Solar System's first 600 Ma (i.e., the "early intense bombardment" of Hartmann et al. 2000). However, the cataclysm is a special case of the LHB and describes when most (or all) of the lunar basins could have formed in a narrow time interval (~3.9 Ga), as noted above.

Based on (1) the impact ages of the Apollo and Luna samples, (2) the assumption that the impact samples collected at specific sites were formed at those sites, and (3) the number of craters counted in an area around those sites, crater size-frequency distributions (CSFDs) are derived as a way to determine the lunar impact flux (e.g., Neukum et al. 1975a, 2001). This method utilizes a count of the number of craters of a given diameter superposed on geologic units (such as impact basins and/or ejecta rays), resulting in cumulative crater counts as a function of diameter and count area, and giving relative ages of that area. Provided samples can be assigned to specific impact craters or basins and thus provide "anchor ages", "absolute model ages" for these regions have been derived and then used to report ages of regions of the Moon from which no impact samples have been collected, including some young crater impact melt sheets (e.g., Hiesinger et al. 2012a) and South Pole-Aitken Basin (e.g., Hiesinger et al. 2012b). Appropriately scaled for size and gravity regime, the Moon's CSFD has even been used to date craters on other planetary bodies (e.g., Wetherill 1975; Hartmann 2005; Werner et al. 2009; Werner and Tanaka, 2011; Marchi and Chapman, 2012; Hiesinger et al. 2016; Lagain et al. 2016; Wulf et al. 2016). This method, however, has recently come under review as we continue to learn more about lunar processes and target properties (e.g., Robbins 2014; van der Bogart et al. 2015; Fassett 2016; Prieur et al. 2016; Miljković et al. 2016).

## 3. Refined Interpretations: Other Lunar Samples, Terrestrial Evidence, and Models

In 2009, Norman noted that "the absolute ages of most lunar basins are effectively unknown" and additionally, that some samples did not have well-established geological context. The National Academy of Sciences in the United States also reported (2007) that since the nature and timing of the lunar impact flux remained unresolved and an important piece of information required for addressing issues in several different science disciplines, interpreting this time-varying impact flux should be a top science priority for NASA's return to the Moon. Thus, perhaps motivated by its probable effect on conditions for life on an early Earth (e.g., Maher and Stevenson, 1988; Sleep et al. 1989), investigators have been looking at a variety of samples (i.e., not just Apollo lunar impact melts), including crystalline melt clasts in meteorites (e.g., Fernandes et al. 2000; Fernandes et al. 2003; Cohen et al. 2005; Fernandes et al. 2013; McLeod et al. 2016; Joy and Arai (2013) provide a nice review), terrestrial and lunar zircons (e.g., Trail et al. 2007, Grange et al. 2011, 2013; Hopkins and Mojzsis, 2015), and lunar impact glasses (e.g., Levine et al. 2005; Delano et al. 2006; Zellner et al. 2009a,b; Zellner and Delano 2015). Additionally, images from high-resolution instruments on lunar orbiters have been used to aid in counting craters (e.g., Neukum et al. 2001) and interpreting geological stratigraphy of the lunar surface (e.g., Fassett et al. 2011, 2012; Spudis et al. 2011; Hiesinger et al. 2012a; Krüger et al. 2016).

**3.1 Lunar Meteorites**

All of the Apollo and Luna samples came from a small portion of the nearside equatorial region of the Moon (see Figure 1) and their impact ages could potentially reflect a sampling bias. Lunar meteorites, on the other hand, have random locations of origin and have been suspected of originating on the lunar farside (Pieters et al. 1983; Ostertag et al. 1985). In fact, impactites in these samples represent mixtures of materials derived from a wide variety of source terrains that are chemically distinct from the regolith at the Apollo and Luna sites and can provide an independent sampling of lunar impacts. Thus, investigations of impactites within lunar meteorites were undertaken in order to elucidate information about the lunar impact flux, and the early lunar impact flux in particular.

For example, in a subset of 30 samples studied, Cohen et al. (2005) found no impact ages older than 4.0 Ga and indicated that one interpretation of these data is that the meteorite impact ages supported the idea of a "cataclysm". However, most studies of the $^{40}Ar/^{39}Ar$ ages of melt clasts in lunar meteorites did not report the age distribution seen in the Apollo lunar samples. Instead, they showed a much larger range of ages, with a broad peak from 2.5 – 4.0 Ga and fewer samples with younger ages (e.g., Fernandes et al. 2000; Cohen *et al.* 2005; Chapman *et al.* 2007; Joy et al. 2011; McLeod et al. 2016).

## 3.2 Terrestrial Samples

As noted earlier, Earth's tectonics, weathering, and rock recycling (among other geological events) have caused the early cratering record to disappear from the terrestrial surface. As a result, lithic remnants older than ~4 Ga on the Earth are rare and siderophile elements (e.g., iridium, platinum) are difficult to find in the oldest sediments on Earth (Ryder et al. 2000). Thus, evidence on Earth for a cataclysmic event around 3.9 Ga is elusive. In a study of the early Archean (>3.7 Ga) terrain of West Greenland, Ryder et al. (2000) searched for Ir signatures, shocked minerals, and remnants of impact ejecta to see if signatures of ancient impacts could be discerned; however, none were found. They attributed this lack of evidence to a fast sedimentation rate and/or a reduced impact flux (Ryder et al. 2000). On the other hand, Anbar et al. (2001) looked at the same metasediments and found elevated (but still low) concentrations of both Ir and Pt; they attributed the low abundances to a sedimentation rate in which accumulating sediments sampled stochastic bombardment by an impactor population governed by a power law mass distribution, such that exogenous Ir and Pt are not usually concentrated in stratigraphic horizons (Anbar et al. 2001).

Though rare, ancient detrital terrestrial zircons may provide evidence for ancient impact events on Earth. Trail et al. (2007) studied a collection of terrestrial zircons and found evidence for heating events in the 3.8 – 4.0 Ga time interval in four of them, while Amelin et al. (1998) found evidence for Pb loss at ~4.1 Ga in others. One interpretation for these observations, since they are similar in assumed age (within uncertainty) to those of a few of the large lunar basins, is that shock heating and metamorphism from the LHB impactors caused fractionation of different elements due to differences in volatilization (Trail et al. 2007). Therefore, these terrestrial zircon samples have been reported to further support the idea of a major short-lived impact event (or series of events) on the Earth (and the Moon). It is noted that at least one zircon with an age of 4.4 Gy supports the existence of a terrestrial crust at this time (Valley et al. 2014).

## 3.3 Dynamical Models

In order for the large lunar basins to form during the purported "cataclysm" in a relatively short period of time centered around 3.85 Ga, some Solar-System wide event is required to disrupt the reservoirs of asteroids and/or comets (i.e., the Asteroid Belt, the Kuiper Belt; e.g., Levinson et al. 2001; Strom et al. 2005; Gomes et al. 2005; Walsh et al. 2012; Werner 2014; Roig et al. 2016). Evidence for the resulting impacts is therefore also seen on other planetary bodies in the inner

Solar System, including Mercury (e.g., Fassett et al. 2011; Werner 2014), Mars (e.g., Frey 2008; Werner 2014), and Vesta and other asteroids (e.g., Bogard 1995; Cohen 2013; Marchi et al. 2013), though the ages of these impact basins and craters are uncertain. Using the interpretation of the lunar impactite ages as representative of the impact cratering history of the inner Solar System, dynamical models of Solar System evolution have been developed to explain this early bombardment. Several scenarios that would lead to a rapid fall-out of material within a relatively short time period have been proposed, including cometary showers and the break-up of a parent body in the Asteroid Belt (e.g., Zahnle and Sleep, 1997). Morbidelli et al. (2001) proposed that the final sweep-up of material during planet formation would have left behind "leftovers" (with a population a few times of that currently in the Asteroid Belt) that would slowly decay on the order of ~60 Ma. The high inclinations (and resulting high velocities) would be enough to create large basin-sized craters on the lunar surface, resulting in a series of impact spikes over time (e.g., Hartmann 1965; Hartmann 1966; Hartmann et al. 2000) but not resulting in any "cataclysm" (Morbidelli et al. 2001). In subsequent papers by different authors, Levinson et al. (2001), Gomes et al. (2005), Tsiganis et al. (2005), and Strom et al. (2005) built upon the Nice (as in Nice, France) model and proposed that the formation and/or migration of the gas giant planets to their current locations (and the swapping of Uranus' and Neptune's positions) may have been the mechanism(s) that disrupted the Asteroid Belt and caused the fall-out of objects in this relatively short period of time. Migration of planets in the inner Solar System has also been proposed (e.g., Walsh et al. 2011, 2012). Even though the resultant Solar System in these models does not exactly resemble the current one (e.g., Nimmo and Korycansky, 2012; Fassett and Minton, 2013), Strom et al. (2005) proposed that impactors since then best resemble the current population of near-Earth asteroids, though this is not universally accepted (Minton et al. 2015; see also Fritz et al. 2014).

## 4. New Data Lead to New Interpretations of the Lunar Impact Flux

New high-resolution orbital data from recent lunar missions, improved resolution and sensitivity of analytical instrumentation, development of new analytical techniques for acquiring ages for lunar samples, re-evaluation of literature data, and updated dynamical models of Solar System evolution that take into account these new observations have led to new interpretations of the early bombardment of the Moon (and by proxy, the Earth). The high-resolution images and data from instruments on the Lunar Reconnaissance Orbiter (LRO) mission, in particular, are changing the way large lunar craters and their ejecta are being identified and characterized and have resulted in a re-examination of the relative ages of the largest and presumed oldest impact basins (i.e., Serenitatis, Tranquilitatis, and Nectaris; Figure 1). Analyses of different lunar and terrestrial impact samples are also improving our understanding of the duration of the early bombardment episodes. With these new data, improved dynamical models are being developed to better explain the observed sample ages and crater populations and to propose the source regions of the impactors. It should be noted that, currently, a major constraint in these models is computational capability that does not permit output resolution similar to that which the samples provide.

## 4.1 Orbital Data

Instruments on board the Lunar Reconnaissance Orbiter (LRO), the Selenological and Engineering Explorer (SELENE), and the Gravity Recovery and Interior Laboratory (GRAIL) spacecraft have provided a wealth of high-resolution data and images that are allowing a closer inspection of the Moon's surface features that have been obscured by aeons of bombardment and volcanism. For example, LRO's Lunar Orbiter Laser Altimeter (LOLA) data provide topographic information that can be used to better define crater (and basin) peak centers and rings, giving better estimates of crater diameters. Frey (2011, 2012, 2015) used this information to find a set of quasi-circular depressions (QCDs) that range in size to >300 km in diameter and speculates that there may be as many as 43-72 previously unidentified impact basins, suggesting more large early impacts. These data, in fact, may show two peaks in impact flux, one around 4.3 Ga and another around 4.05 Ga (Frey 2015). Featherstone et al. (2013) combined both LOLA topography data and gravity data derived from SELENE to infer the presence of 280 possible impact basins, 66 of which were defined as "distinct". These authors, however, do not venture to speculate on the relationship of these 66 basins to the Moon's impact history.

Large crater (D ≥ 20 km) populations can be used to refine and/or constrain the crater size-frequency distribution (CSFD), which has been used as a test of crater/basin ages (see Section 2). Fassett et al. (2012) used this technique with high-resolution LOLA data to re-examine the relative ages of Nectaris and Serenitatis, the latter of which has been thought to be younger (compare ages in 2[nd] and 3[rd] columns of Table 1). More large craters were identified, affecting crater counts (i.e., crater density and CSFD) and the interpretation of the relative emplacement of the respective basin ejecta. The resulting conclusion is that Serenitatis is much older than Nectaris (Fassett et al. 2012), in agreement with a separate study by Spudis et al. (2011). In assessing the high number of craters superposed on Serenitatis' deposits and ring structures, Spudis et al. (2011) surmised that Serenitatis is very old, much older than its assumed 3.89-Ga age (Dalrymple and Ryder, 1996). Moreover, in their study of global wide-angle camera (WAC) images from LRO, Spudis et al. (2011) determined that Imbrium ejecta dominate the nearside. Furthermore, because of the complicated ejecta patterns and sample provenances, the samples collected at the Apollo 17 site and presumed to have originated during the event that formed the Serenitatis basin probably did not (Spudis et al. 2011; but see Hurwitz and Kring, 2016 for a review of the provenances of the impact melt breccias). Spudis et al. (2011), Fassett et al. (2012), and Fernandes et al. (2013) report that Imbrium ejecta (presumably at 3.85 Ga) is littering the nearside and probably contaminating the nearside impact samples, as earlier proposed by Haskin (1998).

Recent data from the GRAIL spacecraft also show large (>300 km diameter) old lunar impact basins (e.g., Neumann et al. 2015), though not as many as identified by Frey (2011, 2012, 2015) or Featherstone et al. (2013). Neumann et al. (2015) reported that anomalies in the gravity measurements of crustal thicknesses in and around the basins' peak rings and rim crests indicate the presence of multiple new basins; additionally six of the known basins were measured to have diameters that are >200 km larger than previously measured. With these newly constrained data, the catalogue of impact basin sizes can potentially be updated, and the current CSFDs recalibrated, possibly affecting the shape of the first billion years of the impact flux curve and thus the relative ages derived for other basins and craters. Neumann et al. (2015) also reported that due to the larger number of >300-km diameter lunar basins now measured on the Moon, a population of impactors with sizes that match those of the current population of asteroids could not have formed them, supporting findings by Minton et al. (2015).

## 4.2 Extraterrestrial Sample Data

Benefitting from new interpretations of old data and more sophisticated analytical techniques resulting in new data, lunar samples are revealing lunar impact ages older than ~3.9 Ga as well as a continuum of impact ages before and after that time. Thus, there is the very interesting development that the "lunar cataclysm" at ~3.9 Ga appears to be falling out of favor in the lunar science community. Instead, a prolonged bombardment, starting as early as 4.2 Ga and lasting until

~3.4 Ga may be a more accurate interpretation of the ages derived from lunar (and other) impact samples ("sawtooth" in Figure 2).

In separate studies, Grange et al. (2010), Liu et al. (2012), Merle et al. (2014), and Mercer et al. (2015) analysed suites of lunar impact samples, including zircons and phosphates in lunar breccia, using recalibrated $^{40}Ar/^{39}Ar$ standards and U-Pb analyses and showed that similar ages were found in many different samples collected from multiple Apollo landing sites. Based on the compositional similarity of these samples to the location of the Imbrium Basin, they concluded that many of these samples were derived from Imbrium (~3.9 Ga) and thus represent one event and not several in a short time period. Grange et al. (2010) further reported that it is possible that the Moon's history was marked by more than one episode of a higher rate of impacts during the first 500 Ma of its history.

It is important to note that as early as 1973, Turner et al. proposed impacts as old as ~4.1 Ga. Moreover, Warner et al. (1977) and Turner and Cadogan (1975) found ~4.2 Ga ages in Apollo 17 breccias, while Maurer et al. (1978) found ages in Apollo 16 breccia older than 4.1 Ga. More recent analyses of other lunar samples are supporting these observations of old (>3.9 Ga) impact ages. For example, Norman et al. (2006) reported an age of ~4.2 Ga in an Apollo 16 breccia, and Fernandes et al. (2013 and references therein) reported 4.2-Gy ages in several Apollo 16 and Apollo 17 impactites. Norman and Nemchin (2014) and Norman et al. (2016) also found evidence for at least one large impact event with a U-Pb age of 4.22 Ga in an Apollo 16 impact breccia, while Fischer-Gödde et al. (2011) determined a Re-Os age of 4.21 Ga in an Apollo 16 impact melt rock. Furthermore, Hopkins and Mojzsis (2015) found evidence in lunar zircons for heating events with U-Pb ages of $4.3 \pm 0.01$, $4.2 \pm 0.01$, and $3.9 \pm 0.01$ Ga. Pb disturbances in lunar zircons have also been attributed to shock events, consistent with impacts older than 3.9 Ga, at ~4.18 Ga (Pidgeon et al. 2006) and 4.1 Ga (Grange et al. 2011). Similarly, meteorites from different parent bodies also suggest several thermal events prior to 4.0 Ga (as compiled in Fritz et al. 2014). Evidence for lunar impacts after ~4.0 Ga to the present is abundant in the lunar samples (e.g., Zellner et al. 2009b, Joy et al. 2011; Fagan et al. 2014; Das et al. 2016) and seen in orbital data (e.g., Speyerer and Robinson, 2014).

## 4.3 Terrestrial Sample Data

Terrestrial samples from Earth's Hadean period are rare and have been subjected to aeons of weathering and tectonism; therefore, interpreting their data is very difficult. However, evidence for old terrestrial Archean impacts has been found in impact structures in South Africa (e.g., Barberton Belt) and Western Australia (e.g., Pilbara Craton). At these sites, multiple impact spherule layers from large distal terrestrial impacts have been determined to have ages between 3.47 and 3.24 Ga (e.g., Lowe et al. 2003; Lowe and Byerly, 2010; Lowe et al. 2015; Glickson et al. 2016), indicating a prolonged history of extraterrestrial bombardment.

The production of these Archean impact spherule beds was modelled by Johnson and Melosh (2012), who reported that the sizes and impact velocities of the asteroids that created these global spherule layers are consistent with a population of impactors that was more abundant 3.5 Ga than it is now. They thus concluded that the impact chronology of these spherule beds is consistent with a gradual decline of the impactor flux after the LHB (Johnson and Melosh, 2012). This conclusion does not, however, provide evidence for the intensity or duration of the bombardment nor does it suggest one way or another the existence of a cataclysm.

## 4.4 Dynamical Models

There is no doubt that impactors bombarded planetary surfaces during the first 600 Ma of Solar System history; therefore some dynamical event (or events) was likely responsible for rearranging the planets, asteroids, and/or comets. However, evidence claiming that all of the large near-side basins formed in a relatively short period of time has weakened in light of new lunar remote sensing data and lunar sample data.

To better match these recent data, as well as the characteristics and configuration of our present-day Solar System (e.g., orbits of the gas giant planets), the Nice Model (as used by Gomes et al. 2005, Tsiganis et al. 2005, and Strom et al. 2005) has been undergoing revisions. In recent studies, Morbidelli et al. (2012) and Bottke et al. (2012) used the framework of the Nice Model to hypothesize the existence of a now-defunct "E Belt" in the Asteroid Belt, between 1.7 and 2.1 astronomical units (AU) from the Sun, in order to explain the presence of the Hungaria family of asteroids. They suggest that a "sawtooth pattern" (Figure 2) can explain today's absence of asteroids in the E-Belt, which was disrupted as the gas giant planets migrated through resonances to their current locations. In this scenario, the large lunar basins were formed over ~400 Ma, starting at ~4.1 Ga, with the Moon (and Earth) continuing to get hit by the E-Belt population until long after 3.7 Ga, with no "cataclysm" at 3.9 Ga. Thus, in addition to explaining the existence of the current high-inclination angle of the Hungaria population of asteroids, this model provides a source of terrestrial impactors (Section 4.3) and supports a long protracted bombardment of the Moon and Earth (e.g., Bottke et al. 2012, Morbidelli et al. 2012; Marchi et al. 2012, 2013; Norman and Bottke, 2017).

## 5. Rethinking the Lunar Impact Flux

### 5.1 Updated Lunar Basin Ages

Given the new evidence from orbital images and data and improved ages of lunar and terrestrial samples, all (or most) of which are supported by updated (but still evolving) dynamical models, a revised timeline for the ages of the lunar impact basins can be presented (3$^{rd}$ column in Table 1). Taking into account the prevalence of near-side Imbrium material that has been found within the Apollo samples (e.g., Haskin 1998; Baldwin 2006; Norman et al. 2010; Spudis et al. 2011), the ages of the largest basins are not well-constrained and have larger errors. It is apparent, however, that their formation is spread over >400 Ma and not in a short time centered ~3.85 Ga. Thus, recent interpretations of orbital data (e.g., Spudis et al. 2011; Fassett et al. 2012) are continuing to revise the relative ages of these basins, possibly calling into question "everything we know about the basin-forming process" (Spudis 2012).

### 5.2 Updated Sample Age Distributions

As mentioned previously, the study of lunar sample ages provides a way to investigate the nature of the impact flux in the Earth-Moon system over time, as no similar archive exists within the terrestrial samples. Smaller lunar regolith samples that were not previously studied due to their size and the lower sensitivity of the equipment 30-40 years ago are contributing more detailed information to the lunar data set. In addition to the lunar meteorites and the lunar impactites, lunar impact glasses are powerful tools that can be used to decipher the time- dependent impact flux in the Earth-Moon system. Formed during the impact itself, lunar impact glasses are pieces of quenched melt that possess the composition of the impact site at which they formed. In one study,

**Table 1.** Comparison of the range of lunar basin ages based on U-Pb and $^{40}$Ar/$^{39}$Ar ages of samples, stratigraphy, and crater counting, as described in the text. Ages in the 2$^{nd}$ column are from Tera et al. (1974), Arvidson et al. (1976), Cadogan and Turner (1976), Drozd *et al.*, (1977), Wilhelms (1987), Ryder (1990), Swindle et al. (1991), Bogard et al. (1994), Dalyrymple and Ryder (1993, 1996), Hartmann (2000), Ryder et al. (2000), Stöffler and Ryder (2001), Baldwin (2006), and Koeberl (2006) and references therein. Ages in the 3$^{rd}$ column are from Norman (2009), Grange et al. (2010), Spudis et al. (2011), Fassett and Minton (2013), Mercer et al. (2015), and Norman and Bottke (2017) and references therein.

| **Crater** | **Age (Ga)** (1974-2006) | **Age (Ga)** (2009-present) |
|:---:|:---:|:---:|
| South Pole - Aitken | $4.05 - \sim4.3$ | $4.0 - 4.4$ (?) |
| Serenitatis | $3.893 \pm 0.009$ | $3.83 - 4.1+$ |
| Nectaris | $3.89 - 3.92$ | $3.92 - 4.2$ (?) |
| Crisium | $3.85 - 3.93$ | $\sim3.9$ (?) |
| Imbrium | $3.85 \pm 0.02$ | $3.72 - 3.93$ |
| Orientale | $3.77 - 3.83$ | $3.72 - 3.93$ |

Delano et al. (2007) identified geochemical signatures for 3.73-Ga glasses from the same landing site and a separate geochemical group of approximately the same age from three different landing sites, supporting the idea of a large global lunar impact event at ~3.73 Ga, which may be the tail end of the heavy bombardment period. Though expected to be numerous, lunar impact glasses older than ~3.8 Ga are rarely found, perhaps as a result of impact gardening which destroys these glasses over time (Zellner and Delano, 2015).

Meteorites that are not derived from the Moon have also shown interesting age distributions that could describe the impact rate in the inner Solar System. For example, impact ages of H chondrite meteorites from the Asteroid Belt have been derived from analyses of argon (Swindle et al. 2009; Wittmann et al. 2010) and U-Th-He (Wasson and Wang, 1991). These age data do not support an impact episode ~3.9 Ga, but they do appear to indicate some other kind of bombardment scenario(s), perhaps unique to the Asteroid Belt, in which there are impacts >3.9 Ga and impacts that taper off until ~3.5 Ga (e.g., Fritz et al. 2014). Impact ages of HED (howardite, eucrite, diogenite) meteorites, presumed to be from Asteroid 4 Vesta, have also been investigated (e.g., Bogard 1995; Bogard 2011; Cohen 2013). Cohen (2013) showed age data that peak around ~3.7 Ga with nothing older than ~3.9 Ga, while Bogard (1995, 2011) noted that some samples indicate a series of argon-system resetting events from 3.4 – 4.1 Gy ago.

With these multiple sample data sets, comparisons between the impact ages obtained from different lunar sample types and the impact ages of other extraterrestrial and terrestrial samples can be made, to see whether the impact rate on the Moon (as represented by the lunar impact samples) is reflected in the impact rate on other planetary bodies. Figure 3a shows a modest data set of lunar impact sample ages and lunar and asteroid meteorite ages plotted on the same scale. Some overlaps among the data sets are obvious, such as the ~500-Ma break-up of the L-chondrite parent body (e.g., Schmitz et al. 2003) in the Asteroid Belt and the impacts between 3.5 Ga and 4.0 Ga, possibly representing the tail end of the LHB. Well-defined ages older than 4.0 Ga and the final sweep up of early Solar-System-forming debris at >4.2 Ga are not seen in the lunar impact glass record nor in the lunar meteorite record. All of the samples, however, appear to indicate a lull in impact activity between ~3.4 Ga and 1.8 Ga. The abundance of impact glass samples with ages <500 Ma may be due to sample preservation (e.g., Levine et al. 2005; Zellner and Delano, 2015).

**5.3 A "Cataclysm" No More**

When taken together, lunar orbital data, terrestrial, lunar, and asteroid sample data, and dynamical modelling of Solar System evolution suggest an extended lunar bombardment from ~4.3 Ga to ~3.5 Ga with evidence for impacts older than ~3.85-3.9 Ga (e.g., Turner et al. 1973; Warner et al. 1977; Norman et al. 2006; Fernandes et al. 2013; Norman et al. 2016) and in contrast to previous reports (e.g., Ryder 1990). Additionally, these sample ages provide evidence for a series of impacts lasting hundreds of millions of years, and not a single "cataclysm" that created all of the large basins on the nearside of the Moon in a short period of time.

**6. Implications for the Origin of Life on Earth**

The nature of the impact history of the Moon, and its presumptive application to Earth, affects our understanding of Earth's habitability and the conditions that existed when life first took hold on a young Earth. In the view of Maher and Stevenson (1988) and Sleep et al. (1989), the Earth was increasingly hostile to life, as the bombardment increased exponentially back in time (i.e., "impact frustration"), though perhaps the heating was only severe locally (e.g., Abramov and Mojzsis 2009; Abramov et al. 2013). In the view of Ryder (2002), the proposed prominent impact spike ~3.9 Ga ago may have had a "benign" effect on life if it originated between 4.4 and 3.85 Ga. In the sawtooth view (e.g., Morbidelli et al. 2012), the bombardment rate was perhaps never exceptionally high. How these impacts would have affected early life is pure speculation, however.

A commonly held view is that the last universal common ancestor (LUCA) may have been thermophilic and/or chemotrophic (e.g., Pace 1991; Stetter, 1996; Boussau et al. 2008; Nisbet and Sleep 2001). This implies that life may have evolved from a LUCA at a time when Earth was hot, due to either volcanism or impacts or a combination of both. Early Archaean high-temperature regimes, such as >100 °C ocean(s), may have been generated by severe impact events (Nisbet and Sleep 2001), though whether or not they were actually sterilizing is not clear (e.g., Zahnle and Sleep, 1997; Ryder 2002; Abramov and Mojzsis 2009). Sterilizing events or not, hyperthermophiles have survived and may have continued to exist through one or more hot-ocean episodes (e.g., Galtier et al. 1999; Nisbet and Sleep 2001).

These hyperthermophiles, though, are not necessarily the first life (Galtier et al. 1999) but may be merely the only kind of organism to survive possible sterilizing events due to impact. Organic compounds, and especially those carrying genetic information, are more likely to survive in low temperatures, which are also necessary for the stability of catalytic polymer configurations. Therefore, it has been proposed that the first biomolecules probably formed during epochs of low temperature (Miller and Lazcano 1995; Bada and Lazcano 2002). A cool early Earth (e.g., Nutman et al. 1997; Mojzsis et al. 2001; Wilde et al. 2001; Watson and Harrison 2005) supports a scenario in which the first living entities appeared and evolved through the RNA world to DNA/protein biochemistry (Bada and Lazcano 2002).

Bada and Lazcano (2002) noted that "if the transition from abiotic chemistry to the first biochemistry on the early Earth indeed took place at low temperatures, it could have occurred during cold, quiescent periods between large, sterilizing impact events". This prolonged bombardment period, with impact episodes separated by hundreds of millions of years (Table 1), rather than a single sterilizing event, is now supported by lunar, terrestrial, and other impact sample ages. Thus, impact frustration and/or sterilization are unlikely, and a cool Earth, hospitable to the development of biomolecules via endogenous and exogenous processes, may have provided the necessary temperature conditions required for biomolecules to be stable. Indeed possible evidence for life very early in Earth's history has been presented: microbial fossils at 3.7 Ga (Nutman et al. 2016); C isotopes at ~3.8 Ga (Schidlowski 1988; Rosing 1999); and putative biogenic carbon at 4.1 Ga (Bell et al. 2015).

Importantly, terrestrial impact samples appear to indicate that the bombardment rate had indeed decreased by 3.4 Ga (e.g., Lowe et al. 2015), around the time that complex life may have been evolving (e.g., Wacey et al. 2011). This life may have been fueled by the delivery of organic material in the form of elemental carbon, hydrogen, oxygen, nitrogen, phosphorous, and sulphur (i.e., CHONPS), amino acids, and sugars (in addition to other biomolecules) by comets, asteroids, and meteorites (e.g., Chyba et al. 1990; Pizzarello et al. 1991, Cooper et al. 2001; Blank et al. 2001; Glavin et al. 2012; McCaffrey et al. 2014; Goesmann et al. 2015). Figure 3b, a modified version of Figure 3a, provides a summary of some of the important terrestrial biological events superposed on the impact rate reflected in the current impact sample age data (with the influence of Imbrium ejecta in the Apollo sample collection de-emphasized in the impact flux ~3.9 Ga). Note that a decline in the impact rate is coincident in time with the first appearances of oxygen (e.g., Bekker et al. 2004; Anbar et al. 2007; Crowe et al. 2013; Satkowski et al. 2015), as well as with the Great Oxidation Event (GOE) and that, in addition to the extinctions at the Cretaceous/Tertiary boundary (Alvarez et al. 1980), more recent impact events may have affected life (e.g., Hedges and Kumar, 2009; Knoll 2014).

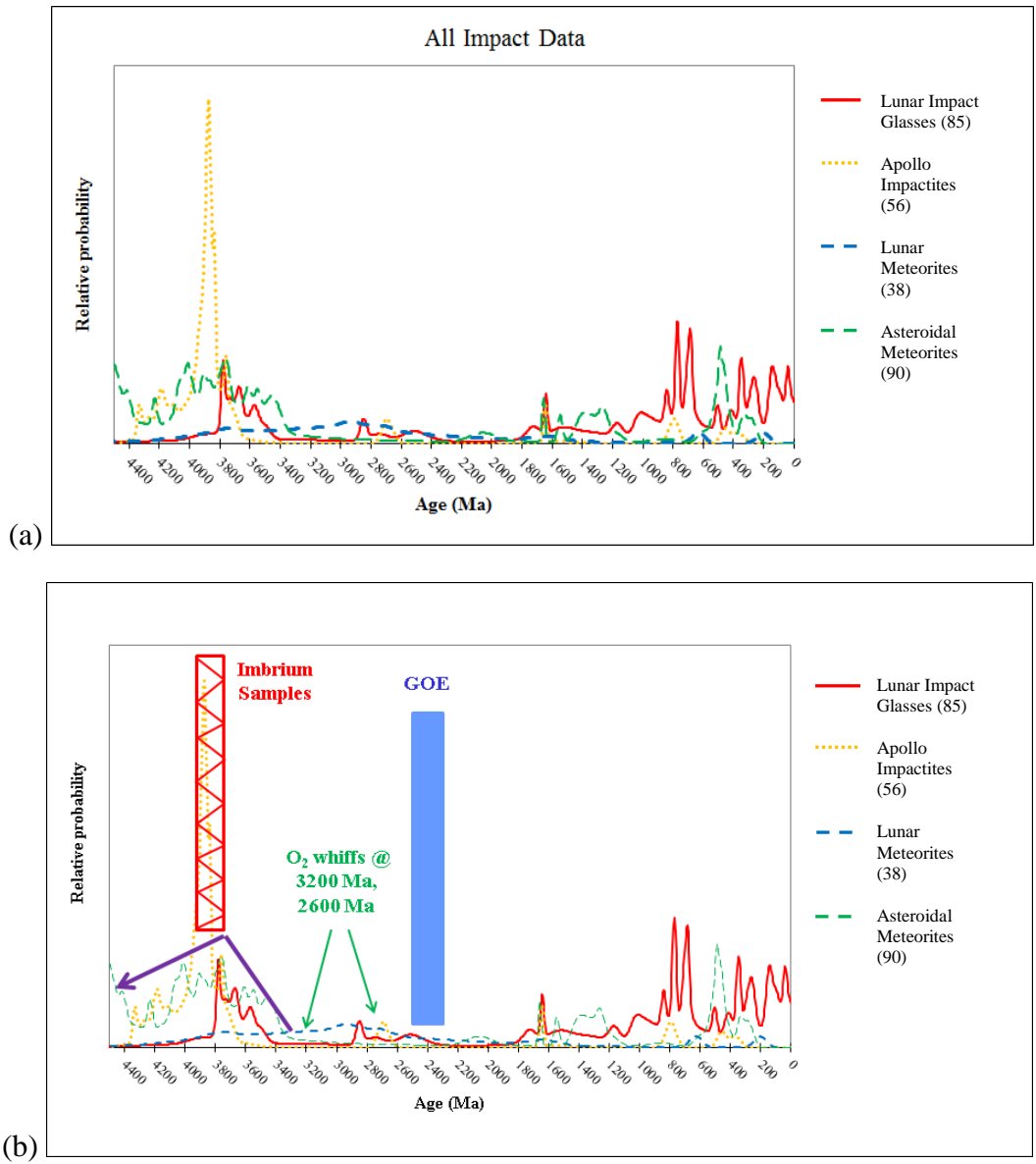

(a)

(b)

**Figure 3.** (a) Relative probability of impact ages occurring in the samples listed. The spike in flux described by the lunar breccias at ~3.9 Ga is most likely due to contamination of Imbrium ejecta that spread to all Apollo landing sites, as described in the text. (b) Important terrestrial biological events shown along with impact flux scenarios as represented by impact sample ages. The influence of Imbrium ejecta in the Apollo sample collection (~3.9 Ga) has been de-emphasized, and evidence for oxygen, including the GOE, is noted. References for the lunar sample age data in both (a) and (b) are from Compston et al. (1972), Papastassiou and Wasserburg (1972), Eberhardt et al. (1973), Mark et al. (1974), Turner and Cadogan (1975), Cadogan and Turner (1976), McKay et al. (1978), Spangler et al. (1984), Reimold et al. (1985), Borchardt et al. (1986), Bogard et al. (1991), Dalrymple and Ryder (1993, 1996), Bogard (1995), Ryder et al. (1996), Stöffler and Ryder (2001), Cohen et al. (2005), Norman et al. (2006), Cohen et al. (2007), Delano et al. (2007), Hudgins et al. (2008), Nemchin and Pidgeon (2008), Fernandes et al. (2000), Zellner et al. (2009a,b), Hui (2011), Cohen (2013), Fernandes et al. (2013), and Zellner and Delano (2015). Terrestrial oxygen data in (b) are from Bekker et al. (2004), Anbar et al. (2007), Crowe et al. (2013), and Satkowski et al. (2015).

## 7. Conclusion

Due to the fact that the Moon has a long history of almost no geologic activity and therefore preserves its long record of impacts, understanding the rate of impacts on the Moon allows us to draw conclusions about the impact rate on Earth (since they are so close together in space). With this information, a better estimate for when and how soon conditions on Earth became suitable for life as we know it can be determined. The Moon's ancient cratered lunar highlands provide evidence that the Moon, and most likely the inner Solar System, was heavily bombarded in its first billion years (e.g., Hartmann 1965; Stöffler and Ryder, 2001; Neukum et al. 2001), but a scenario in which all (or most) of the large nearside basins formed in a "cataclysm" is no longer substantiated by recent data and re-evaluation of literature data. It does appear that some dynamical event was responsible for material bombarding the planetary surfaces, but orbital and sample evidence suggests that a cataclysm, and especially the extreme case of the cataclysm (Ryder, 1990), is unlikely. Instead, a protracted period of impact events lasted for at least 400 Ma (e.g., Morbidelli et al. 2012; Bottke et al. 2012; Fritz et al. 2014). Recent high-resolution orbital data and images (most notably from LRO and GRAIL), more refined techniques for studying small lunar, terrestrial, and other impact samples and a better understanding of their ages, and improved dynamical models based on orbital and sample data have caused a paradigm shift in how we think about the lunar impact rate and how it applies to Earth. The long-held idea of a "lunar cataclysm" at ~3.9 Ga is being replaced by the idea of an extended lunar bombardment from ~4.2 Ga to 3.5 Ga. While the effects of this prolonged bombardment on the evolution and development of life on Earth have yet to be closely investigated, the timeline for the assembly of the first biomolecules has been lengthened.

## Acknowledgements

NEBZ acknowledges support from the NASA LASER and Solar System Workings programs (NNX11AB28G, NNX15AL41G), the NSF Astronomy and Astrophysics program (#1008819, #1516884), and the Hewlett-Mellon Fund for Faculty Development at Albion College. NEBZ sincerely thanks John W. Delano for always insightful conversations related to our holistic approach to interpreting lunar impact glass data. A thorough review by Vera Fernandes and comments from an anonymous reviewer helped to improve this manuscript. NEBZ also thanks

Marc Norman and Vanessa McCaffrey for thoughtful comments on earlier drafts of this manuscript and Alan Schwartz for editorial handling.